\documentclass[twocolumn]{article}
\usepackage{graphicx}   
\usepackage{subfigure}  
\usepackage[left=2.0cm,top=2.5cm,right=2.5cm,bottom=2.5cm,marginratio={1:1},scale=10]{geometry} 
\usepackage{multicol}   
\usepackage{changepage} 
\usepackage{authblk}    
\usepackage{amsmath}    
\usepackage{tabularx}   
\usepackage{abstract}   
\usepackage{amsfonts}   
\usepackage{url}        
\usepackage{etoolbox}
\patchcmd{\thebibliography}{\section*{\refname}}{}{}{}

\usepackage{titlesec} 
\renewcommand\thesection{\Roman{section}} 
\renewcommand\thesubsection{\Roman{subsection}} 
\titleformat{\section}[block]{\centering}{\textbf\thesection.}{0.2em}{}
\titleformat{\subsection}[block]{\large}{\thesubsection.}{0.2em}{}

\usepackage{lipsum}
\let\OLDthebibliography\thebibliography
\renewcommand\thebibliography[1]{
\OLDthebibliography{#1}
\setlength{\parskip}{0pt}
\setlength{\itemsep}{0pt plus 0.3ex}}

\usepackage[font=footnotesize ]{caption}              

\begin{document}
\twocolumn[
\begin{@twocolumnfalse}
\vspace{-8ex}
\title{\fontsize{13pt}{13pt}\selectfont\textbf{Classical Propagation in the Quantum Inverted Oscillator}}
\author{\fontsize{10pt}{10pt} \selectfont{Carla M. Q. Flores}\thanks{carla.mariela729@gmail.com} }
\affil{\fontsize{10pt}{10pt} \emph{Department of Physics, Mayor de San Andr\'es University, La Paz, Cota-Cota 27, Bolivia}\\
(Dated: December 5, 2016) \vspace{-8ex}}
\date{}
\maketitle
\begin{abstract}
\begin{adjustwidth*}{0.5cm}{0.5cm} 
We emphasize the fact the evolution of quantum states in the inverted oscillator (IO) is reduced to classical equations of motion, stressing that the corresponding tunnelling and reflexion coefficients addressed in the literature are calculated by considering only classically trajectories. The Wigner function formalism is employed to describe the IO classical dynamics, subsequently leading to the introduction of the Ambiguity function lying in the so-called Reciprocal phase space. Our findings, show that the Ambiguity function behavior, subjected to the IO, allude a classical propagation with an associated integral of motion, and complex conjugated doubly degenerate energy states. 
\\
\\
PACS numbers: 07.05.Kf 
\vspace{4ex}
\end{adjustwidth*}
\end{abstract}
\end{@twocolumnfalse}]
\saythanks 
\section{\small{\textbf{INTRODUCTION}} }
The inverted oscillator (IO) is one of the few completely solvable physical systems in both quantum  
and classical mechanics. Its classical Newtonian solutions are expressed in terms of hyperbolic functions 
that diverge exponentially in time, while the quantum counterpart leads to continuous, doubly degenerate energy eigenstates with no ground state defined. Ever since Barton's thesis \cite{barton1986quantum}, the IO has been studied with 
high interest motivated by several technological applications and theoretical developments such as fission 
dynamics \cite{PhysRevC.56.1025}, string theory \cite{cremonini2005tachyon} and universe models \cite{PhysRevD.32.1899,PhysRevD.66.023507}.

The evolution of the IO is acknowledged in the literature as being classical; we revisit this subject  
under the light of a free-coordinate formulation of the time-evolution operator in the \emph{Hilbert phase space}, introduced in the recent publications \cite{bondar2012operational,PhysRevA.88.052108,PhysRevA.92.042122}, that naturally allows to introduce the system's features in the phase space and in the reciprocal phase space. The paper is structured as follows: In section 2, we review the problem from the point of view of Newtonian mechanics by describing the 
classical phase portrait. Quantum mechanics in the \emph{Hilbert phase space} \cite{bondar2012operational,PhysRevA.88.052108,PhysRevA.92.042122} is briefly outlined in section 3, with the aim to demonstrate the well-known equivalence between classical and 
quantum evolution under quadratic Hamiltonians. In addition, Wigner function is introduced due to its particularly hallmark of being helpful to gain insight in the role of both quantum and classical mechanics. In section 4, we bring up for discussion the controversy of the quantum 
tunneling coefficient associated to the IO as described in the literature \cite{balazs1990wigner,heim2013tunneling,guo2011quantum}. In section 5, the IO classical and quantum reciprocal phase spaces are studied for the first time, revealing an additional integral of motion. Finally, in the last section we provide the conclusions.

\begin{figure}[pt!]
\begin{tabular}{cc}
\includegraphics[scale=0.51,angle=0]{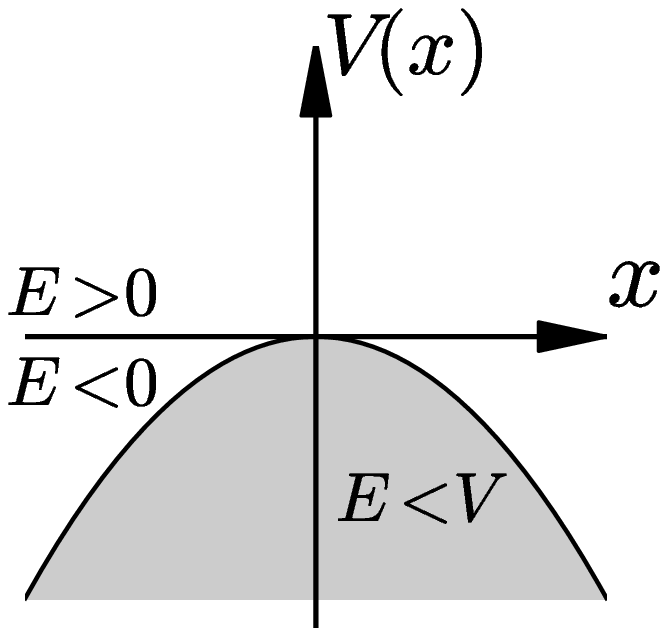} &
\includegraphics[scale=0.75,angle=0]{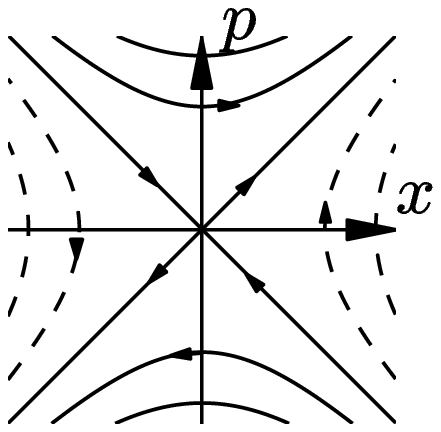} \\
(a) & (b)\\
\end{tabular}
\caption{(a) Inverted oscillator potential barrier. (b) Classical phase$ $-$ $space portrait: solid or dashed lines 
correspond to particles with positive or negative energies respectively. The direction
of motion is represented by the arrows. }
\label{figure-1}
\end{figure}
\section{\small{\textbf{NEWTONIAN PICTURE}}}
\begin{figure*}[t!]
\centering
\hspace{0em}\includegraphics[scale=0.8,angle=0]{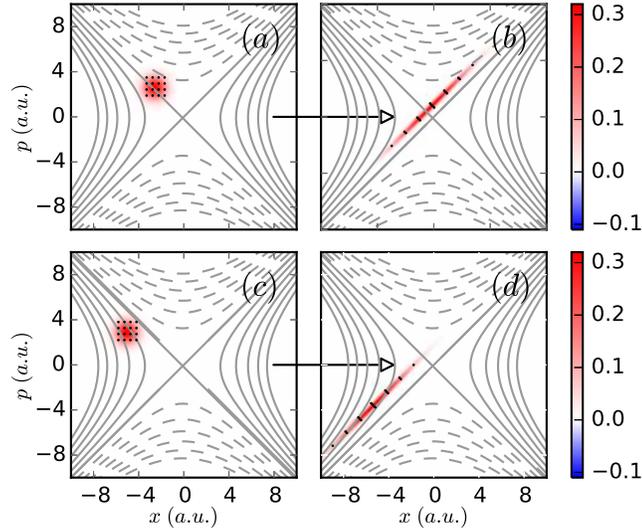}
\caption{Gaussian Wigner functions $ W(x,p)$ of energies $\mathbf{E}_{1}=-0.5 $ [(a)\&(b)], and $\mathbf{E}_{2}=-8$ [(c)\&(d)] subjected to the IO at times $t_{0}=0 \ a.u. $ and  $t=1.5 \ a.u.$ ($\hbar=\omega=1 a.u.$). Black dots depict classical particles moving under Newton's laws and solid lines display the level set of classical the Hamiltonian Eq.(1).}
\label{figure-2}
\end{figure*}
The classical inverted harmonic oscillator is characterized by the classical Hamiltonian  
\begin{align}
\label{class_h-Eq}
\mathcal{H}(x,p) = \frac{p^{2}}{2m} - \frac{1}{2}m\omega ^{2}x^{2}=E, %
\end{align}
where $x$ and $p$ are the canonical position and momentum variables, $m$ is the particle{}'s mass, $\omega$ denotes the repulsion parameter, and $E$
stands for the energy. This model is completely integrable and shows non periodic behavior
\begin{equation}
\label{newton_sol-Eq}
  \left\lbrace
  \begin{array}{l}
     x(t) = x_{0}\cosh(\omega t) + p_{0}\sinh(\omega t) / m \omega \\
     p(t) = m\omega x_{0} \sinh(\omega t) + p_{0}\cosh(\omega t). \\
  \end{array}
  \right.
\end{equation}
The classical phase space pictured by Fig. \ref{figure-1}(b) displays asymptotic lines emerging from the origin
called separatrices $ p=\pm m\omega x $, and sets of hyperbolas around the saddle point $(x=0,p=0)$. The separatrices portray phase space trajectories of zero energy $E=0$, and divide the phase space into four quadrants: the \emph{upper} \& \emph{lower} sets of hyperbolic lines that represent phase space trajectories of particles with positive energies $E>0$, moving over the barrier; and the \emph{right} \& \emph{left} sets correspond to phase space trayectories of particles of negative 
energies $E<0$, reflected from the barrier.

The approach towards the saddle point demands infinite time. It can only be performed by particles settle 
down on the separatrix  $ p=-m\omega x $, often referred as the \emph{stable  separatrix}, according to 
\begin{equation}
 x(t)= x_{0}e^{-\omega t}, %
\end{equation}
and the \emph{unstable  separatrix } $p=m\omega x$, describes particles moving away the saddle 
point 
\begin{equation}
 x(t)= x_{0}e^{\omega t}. %
\end{equation}
\section{ \small{\textbf{CLASSICAL PROPAGATION OF WAVE PACKETS UNDER QUADRATIC HAMILTONIANS}}}
It is well-known that quantum dynamics of quadratic Hamiltonians can be exactly reduced to  
classical equations of motion. In this section, we arrive to the same conclusion by exact algebraic manipulation of the quantum equations of motion in the \emph{Hilbert phase space}.

Consider the abstract form of the von Neumann equation in the \emph{Hilbert phase space} \cite{bondar2012operational,PhysRevA.88.052108,PhysRevA.92.042122}
\begin{equation}
\label{vonNeumann-Eq}
i\hbar\frac{ d }{ d t}\left|\rho(t)\right\rangle 
=[H(\hat{\boldsymbol{x} } , \hat{ \boldsymbol{p}}) - H(\hat{\boldsymbol{x}}' , \hat{\boldsymbol{p}}')]\left|\rho(t)\right\rangle. %
\end{equation}
where
\begin{align}
\label{standart_commuting_relations}
[\hat{\boldsymbol{x} } , \hat{\boldsymbol{p} }  ] = i \hbar, \quad [ \hat{\boldsymbol{x} }' , \hat{\boldsymbol{p} }'  ] = -i \hbar, 
\end{align}
while the commutator of the cross-terms vanish $[ \hat{\boldsymbol{x} } , \hat{\boldsymbol{x} }'  ] = [ \hat{\boldsymbol{x} } , \hat{\boldsymbol{p}}'  ] = [ \hat{\boldsymbol{p}}' , \hat{\boldsymbol{p} }  ] =[ \hat{\boldsymbol{p}}' , \hat{\boldsymbol{x} }  ] = 0$. 
The set of position and momentum operators 
$(\hat{\boldsymbol{x}},\hat{\boldsymbol{p}},\hat{\boldsymbol{x}}', \hat{\boldsymbol{p}}')$ 
are rewritten in terms of a new set of operators $(\hat{x}, \hat{p}, \hat{\lambda}, \hat{\theta})$, called the extended four-operator algebra, through Bopp transformations \cite{Bopp1956}
\begin{align}
  \label{eq-x-1-b} 
 \hat{\boldsymbol{x} }  = \hat{x}-\frac{\hbar}{2}\hat{\theta}, \quad \hat{\boldsymbol{x} }' = \hat{x}+\frac{\hbar}{2}\hat{\theta},\\
  \label{eq-x-2-b}
  \hat{\boldsymbol{p}} =\hat{p}+\frac{\hbar}{2}\hat{\lambda} , \quad \hat{\boldsymbol{p}}'=\hat{p}-\frac{\hbar}{2}\hat{\lambda}.
\end{align}
Commutators relations of the operators $(\hat{x}, \hat{p}, \hat{\lambda}, \hat{\theta})$ \cite{bondar2012operational} are constructed such that  Eqs. (\ref{eq-x-1-b}) and (\ref{eq-x-2-b}) attain the standart commuting relations given in Eq. (\ref{standart_commuting_relations})   
\begin{align}
\label{extended-commutators}
[\hat{x},\hat{p}]=0,   \quad  [\hat{x},\hat{\lambda}]=i,   \quad [\hat{p},\hat{\theta}]=i,  \quad [\hat{\lambda},\hat{\theta}]=0.
\end{align}
Then, the substitution of Eqs. (\ref{eq-x-1-b}) and (\ref{eq-x-2-b}) in Eq. (\ref{vonNeumann-Eq}), enable to write von Neunmann equation in the
\emph{Hilbert phase space}
\begin{multline}
\label{vonNeumann-Eq2}
i\hbar\frac{d }{d t}\left|\rho(t)\right\rangle 
 = \Big[ H(\hat{x}-\frac{\hbar}{2}\hat{\theta} , \hat{p}+\frac{\hbar}{2}\hat{\lambda}) \\
- H(\hat{x}+\frac{\hbar}{2}\hat{\theta} ,\hat{p}-\frac{\hbar}{2}\hat{\lambda})  \Big]\left|\rho(t)\right\rangle. %
\end{multline}
Note the quantum state is represented by the ket $\left|\rho(t)\right\rangle$, instead of the density state operator
$\hat{\rho}$; and is subjected to specific coordinate representations realized by the projections on either of the four conceivable ket representations, parametrized by the eigenvalues of a pair of commuting operators belonging to the set $( \hat{x}, \hat{p}, \hat{\lambda},\hat{\theta})$ \cite{PhysRevA.88.052108,PhysRevA.92.042122}
\begin{align}
x-p, \qquad x-\theta, \qquad \lambda-p, \qquad \lambda -\theta. 
\end{align} 
For instance, in the phase space representation [$x$-$p$] the ket is realized as $ \left\langle xp\right|\rho(t)\rangle$, and the four-operator algebra is accomplished by 
\begin{align}
\label{xp-diff-op}
\hat{x}=x, \quad \hat{p}=p, \quad 
\hat{\lambda}=-i\frac{\partial} {\partial x}, \quad  \hat{\theta} =-i \frac{\partial}{\partial p}.
\end{align}
[See Appendix A for more details.].

The classical limit of von Neumann equation yields 
\begin{multline}
\label{KvN-Eq}
i \frac{d }{d t}\left|\Psi(t)\right\rangle  = 
 \Big[ \frac{\partial }{\partial \hat{p}} H(\hat{x}, \hat{p}) \hat{\lambda}  - 
\frac{\partial }{\partial \hat{x}} H(\hat{x}, \hat{p}) \hat{\theta} 
 \Big]\left|\Psi(t)\right\rangle, %
\end{multline}
which was properly identified in Ref. \cite{PhysRevA.88.052108} as the classical Koopman-von Neumann (KvN) equation \cite{koopman1931hamiltonian,mauro2002koopman,mauro2003topics,deotto2003hilbert,PhysRevA.88.052108}, rather than the classical Liouville's 
equation. It is particularly represented in the phase space [$x$-$p$] by 
\begin{multline}
  \frac{\partial }{\partial t} \Psi(x,p;t) =
 \frac{\partial }{\partial x} H( x, p)  \frac{\partial }{\partial p}\Psi (x,p;t) \\
 -\frac{\partial }{\partial p} H( x, p) \frac{\partial }{\partial x}\Psi (x,p;t),
\label{KvN-Eq-xp}
\end{multline}
where $ \Psi (x,p;t) = \left\langle xp\right|\Psi(t)\rangle $ is the classical Koopman–von Neumann 
wave function connected with the classical Liouvillian probability density $\rho(x,p;t)$ through $\rho(x,p;t) = {\left |\Psi(x,p;t)  \right |}^2$. The classical Liouville's equation is recovered by utilizing the chain rule in the later relation and Eq.(\ref{KvN-Eq-xp})     
\begin{multline}
\frac{\partial }{\partial t} \rho (x,p;t) =
\frac{\partial }{\partial x} H( x, p)  \frac{\partial }{\partial p} \rho(x,p;t) \\
-\frac{\partial }{\partial p} H( x, p) \frac{\partial }{\partial x} \rho(x,p;t). 
\label{Liouville-xp}
\end{multline}

Regarding quadratic hamiltonians, the below general hamiltonian is proposed 
\begin{align}
 H( \hat{\boldsymbol{x}}, \hat{\boldsymbol{p}}  ) 
= \frac{\hat{\boldsymbol {p} }^2}{2m}  - a +b\hat{\boldsymbol{x}}+c\hat{ \boldsymbol{x} }^{2},
\end{align}
where $a, b,$ and $c$ are constant coefficients, and $\hat{\boldsymbol {x}}$, $\hat{\boldsymbol {p}}$ are the standard position and momentum operators. Then, by means of Eq. (\ref{vonNeumann-Eq2}) the related von Neumann equation in the \emph{Hilbert phase space} is obtained 
\begin{align}
\label{Moyal_polyn-Eq}
i\frac{d }{d t}\left|\rho(t)\right\rangle =\Big[\frac{\hat{p} \hat{\lambda}}{m} +(b+2c\hat{x})\hat{\theta}\Big]\left|\rho(t)\right\rangle, 
\end{align}
more precisely, for a certain quadratic Hamiltonian von Neumann equation is brought to the form \cite{bondar2012operational}
\begin{align}
\label{Moyal_quadratic2-Eq}
i\frac{d }{d t}\left|\rho(t)\right\rangle
 =\Big[\frac{\partial }{\partial \hat{p}} H(\hat{x}, \hat{p}) \hat{\lambda} - \frac{\partial }{\partial \hat{x}} H(\hat{x}, \hat{p}) \hat{\theta} 
 \Big]\left|\rho(t)\right\rangle. %
\end{align}
Thus equations (\ref{KvN-Eq}), (\ref{Moyal_polyn-Eq}), and (\ref{Moyal_quadratic2-Eq}) are quantum compliant due to $\hbar$ is inherently cancelled without taking the classical limit $\hbar \rightarrow 0$. So, quantum and classical \emph{evolution} is identical as long as quadratic Hamiltonians are considered, however, be aware that the quantum state $\left|\rho(t)\right\rangle$ among other quantum restrictions obeys the uncertainty principle while the classical Koopman–von Neumann wave function $\left|\Psi(t)\right\rangle$ can be more arbitary since classical states develop eventually higher and higher resolution without limit \cite{zurek2001sub}. 

On the other hand, the phase space representation [$x$-$p$] of the ket $ \left\langle xp\right|\rho(t)\rangle$ 
is proportional to the Wigner function \cite{wigner1932quantum,PhysRevA.88.052108} [See Appendix B for more details.]
\begin{align}
\label{Wigner-function-Eq}
  W(x,p;t) =  \frac{1}{\sqrt{2\pi \hbar}} \langle  xp | \rho (t)  \rangle.
\end{align}
The equation of motion for the Wigner function is known as
Moyal's equation \cite{moyal1949quantum,zachos2005quantum,curtright2012quantum,PhysRevA.92.042122}, 
and coincide with the classical Koopman-Von Neumann equation for quadratic Hamiltonians. Hereafter, within this environment the quantum inverted oscillator (IO) is treated 
\begin{align}
\label{IO-hamil-Eq}
H = \frac{\hat{\boldsymbol {p} }^2}{2m}  -\frac{1}{2} m \omega ^{2} \hat{ \boldsymbol{x} }^{2},
\end{align}
where $m$ accounts the particle's mass, $\omega$ is the repulsion parameter, and $\hat{\boldsymbol {x}}$, $\hat{\boldsymbol {p}}$ are the position and momentum operators. Employing Eqs. (\ref{Moyal_polyn-Eq}) and (\ref{xp-diff-op}), the IO Moyal's equation is attained  
\begin{align}
\label{Moyal_IHO-Eq}
-\frac{\partial W(x,p;t)}{\partial t} =\left[ \frac{p}{m} \frac{\partial}{\partial x} + m {\omega}^2 x \frac{\partial}{\partial p} \right]W(x,p;t). %
\end{align}
Numerical propagation of this equation is carried out utilizing Pure Gaussian Wigner functions, as initial states 
\begin{align}
\label{W_init-Eq}
W(x,p,t_{0}=0) 
= \frac{1}{\pi \hbar}e^{- (m \omega^2 (x-x_{0})^{2} + \frac{(p-p_{0})^{2}}{m } ) /(\hbar\omega) }.%
\end{align}
Purity condition, Eq.(\ref{purity-Eq}), stipulates that $W(x,p;t)$ might be faithfully represented in terms of a Schr\"odinger'\ s wavefunction up to a global phase factor. 
\begin{align}
\label{purity-Eq}
2\pi \hbar \int W^{2}(x,p)dxdp=1. %
\end{align}

For illustrative purposes, the numerical propagation of two Wigner functions 
of energies $\mathbf{E}_{1}=-0.5$ and $\mathbf{E}_{2}=-8$  were implemented using the spectral Split-operator 
method \cite{PhysRevA.92.042122} (See Python code in \cite{WignerIO}); screenshots of the evolution are displayed in Fig.(\ref{figure-2}). The studied Wigner functions move along the classical phase space trajectories following the level sets 
of the classical Hamiltonian Eq.(\ref{class_h-Eq}), rewarded by the fact that Moyal's equation and the classical Koopman von Neumann equation are identical for quadratic Hamiltonians. Thus, proceed to the comparison with the classical phase-space portrait is completely natural, for example a state that approaches the barrier from the left or right side is located above or below the unstable separatrix, and according to the arriving direction the positive energy components of Wigner function might be located in the upper or lower quadrants (Positive energy trajectories of the classical phase portrait), the zero energy components are settled down over the stable separatrix, and the  negative energy components might be placed on the right or left quadrants (Negative energy trajectories of the classical phase portrait). More importantly, the positive, zero and negative energy components 
of the Wigner function weights the particle's contribution: (i) to move over the barrier, (ii) to stop at the top,
or (iii) to be reflected. Another theoretical argument that support the classically evolution description 
is that the positive-definite Gaussian Wigner functions, set as initial states, remain positive distributions throughout the evolution generated by Eq.(\ref{Moyal_IHO-Eq}). This is ensured by the fact that for pure states, Gaussians are the only possible 
positive Wigner functions, according to Hudson's theorem \cite{hudson1974wigner}. As a result, the evolution of the Wigner function under quadratic Hamiltonians might be well sketched out by Newtonian particles. 

The previous arguments and simulations prove that the evolution of the Wigner function under the IO is effectively classical 
in the sense that the equation of motion is free from $\hbar$. Nevertheless, Planck's constant still enters as a parameter
in the initial state ensuring that the state is consistent with the uncertainty principle among others 
quantum conditions \cite{ganguli1998quantum,tatarskiui1983wigner}, which for quantum pure states remains valid all along the propagation.  
Complementarily, it is noteworthy mention that the Epistemically Restricted Liouville mechanics \cite{bartlett2012reconstruction,jennings2015no} is able to reproduce many quantum phenomena of Gaussian Quantum mechanics \cite{olivares2012quantum,weedbrook2012gaussian} by emulating the uncertainty principle on the canonical variables and setting up the maximum entropy principle. However, the scope of this classical treatment was recently investigated in Ref. \cite{ahmadzadegan2016classicality} by couplying a classical oscillator with a gaussian quantum oscillator, both equivalent under this criteria; the evolution showed that the quantum sector of the former violates the uncertainty principle stating that the quantum features cannot be completely overshadowed.
\section{ \small{\textbf{APPROACH TOWARDS THE IO BARRIER }}}
Quantum tunneling is a fundamental quantum mechanical effect where a particle penetrates a potential barrier energetically higher than the particle's total energy, entering in the \emph{classically forbidden region}, thus, leading a measurable probability of crossing the other side of the barrier, otherwise prohibited by the classical mechanics. 

For states approaching the IO barrier from the right (left) side, shown in Fig.~\ref{figure-1}(a), the prohibited regions are displayed in the classical phase space portrait, Fig.~\ref{figure-1}(b), laying within the lower-half portion of the left quadrant (the upper-half portion of the right quadrant). Notwithstanding, analyzing the IO energy eigenstates, the authors in Ref. \cite{balazs1990wigner} derived an analytic 
expression for the tunneling coefficient $T$ considering only classically allowed phase space trajectories, corresponding to 
positive energy components of Wigner function, located above the top of the barrier. This result is in contradiction with the 
conventional WKB theory where this effect comes from the use of complex trajectories, nevertheless,
it was justified by the presence of separatrices in the classical phase space. The unstable separatrix  
automatically prohibits the flow of Wigner functions
across the forbidden regions while the stable separatrix spreads the Wigner functions into two separated branches  
of positive and negative energy components, whenever the states cross it, turning the Wigner function strongly non local, and since the interference forms the basis for the semiclassical evaluation, this leads to write the semiclassical approximation of the Wigner function as WKB waves undergoing interference. On this way, for states with total \emph{negative energy}, Fig. \ref{figure-2}, the positive energy components of the Wigner function were associated with the tunneling coefficient $T$,  and the negative energy components with the reflection coefficient $R$. Nonetheless, further research in Refs. \cite{maitra1997barrier,maitra1997tunneling} extended the study of 
the conventional semiclassical approach by using  the path integral framework in both the time and 
energy domains. They found that for a 
complete and accurate semiclassical approximation of the quantum propagator applied to  
tunneling problems, the semiclassical propagator must consider two contributions from: 
(i) above the barrier trajectories and (ii) below the barrier trajectories associated 
with the tunneling loops. Ultimatelly, demonstrating that the later contribution becomes dominant at long times and far endpoints in barriers that flatten out at large distances  $\lim_{x\rightarrow \infty }V(x)=0$, regarding
the calculation of the tunneling coefficient. Moreover, as an example of barriers that do not flatten out at large distances, 
they studied the IO, for which the second 
contribution to the semiclassical propagator vanishes at all, implying that no tunneling trajectories will develop, since the
semiclassical propagator is exact and only picks up classically allowed trajectories. 
In the light of these insights, they refute the findings of Ref. \cite{balazs1990wigner}, however, more recently papers about the IO \cite{heim2013tunneling,guo2011quantum}
deal with tunneling and reflection coefficients $T$ and $R$ as a result of trajectories in the classical phase space 
that exactly draw the time-evolution of the Wigner function.
\section{\small{\textbf{RECIPROCAL PHASE SPACE}}}
Appealing to the Ambiguity function $A(\lambda,\theta)$ \cite{cohen1985generalized,cohen1989time,PhysRevA.92.042122} the inverted oscillator dynamics is alternatively reformulated in the $\lambda-\theta$ representation, hereafter referred as the \emph{Reciprocal phase space}, bearing that $A(\lambda,\theta)$ is obtained through a two dimensional Fourier transform on the Wigner function 
\begin{align}
\label{A-eq-1}
A(\lambda,\theta)= \int W(x, p) e^{-i(\lambda x + p \theta)}dx dp.
\end{align}
The motion equation for the Ambiguity function rewritten for the IO hamiltonian, is read as [See Appendix C for details.]
\begin{align}
\label{Motion_Ambiguity_IO-Eq}
\frac{\partial A(\lambda,\theta;t)}{\partial t} =
\left[ \frac{\lambda}{m} \frac{\partial}{\partial \theta} + m {\omega}^2 \theta \frac{\partial}{\partial \lambda} \right]A(\lambda,\theta;t),
\end{align}
where the characteristics of 
this partial differential equation provides the ensuing reciprocal classical phase-space trajectories
\begin{align}
\label{R_newton_sol-Eq}
  \left\lbrace
  \begin{array}{l}
     \lambda(t) = \lambda_{0}\cosh(\omega t) - m \omega \theta_{0}\sinh(\omega t) \\
     \theta(t) = -\lambda_{0}\sinh(\omega t)/m\omega + \theta_{0}\cosh(\omega t). \\
  \end{array}
  \right.
\end{align}
It turns out that this system obeys the ordinary differential equations below
\begin{align}
\label{R_canonical-Eq}
\frac{d \lambda}{dt} =  \frac{\partial {\mathbb{H}}(\lambda,\theta) }{\partial \theta}, \quad
\frac{d \theta}{dt} = -\frac{\partial {\mathbb{H}}(\lambda,\theta) }{\partial \lambda}.
\end{align}
where ${\mathbb{H}}(\lambda,\theta)$ is a function constructed in the reciprocal phase space, such that 
${\mathbb{H}}(\lambda,\theta) = {\cal E}$
\begin{align}
\label{R_Hamiltonian-Eq}
{\mathbb{H}}(\lambda,\theta)=-\frac{1}{2}m \omega^2 \theta^2 + \frac{\lambda^2}{2m} = {\cal T}(\theta)+
{\cal V}(\lambda)= {\cal E}, %
\end{align}
\begin{align}
 {\cal T}(\theta)  =  -\frac{1}{2}m \omega^2 \theta^2, \quad
 {\cal V}(\lambda) =  \frac{\lambda^2}{2m},
\end{align}
and ${\cal T}(\theta)$ is a scalar function related to the motion, 
whereas ${\cal V}(\lambda)$ plays the analogue role of the barrier. 

In comparison to the classical phase space, the reciprocal phase space Fig.(3-b), also exhibits the stable and the unstable separatrices for particles of ${\cal E}=0$, described by the next asymptotes and displacement rules 
\begin{align}
\label{R_Aproximation-Eq}
\
\theta = \frac{\lambda}{m \omega},  \quad \lambda(t)=\lambda_{0}e^{-\omega t},%
\
\end{align}
\begin{align}
\label{R_Away-Eq}
\
\theta = -\frac{\lambda}{m \omega},  \quad \lambda(t)=\lambda_{0}e^{\omega t}.%
\
\end{align}
It follows that particles arriving the $\cal V(\lambda)$ barrier from the left (right) side are portrayed down (up) the unstable separatrix $\theta = -\lambda/m \omega$, besides, the upper and lower sets of hyperbolas describe particles of ${\cal E}<0$ that pass below the $\cal V(\lambda)$ while the right and left sets represent particles of ${\cal E}>0$ reflected from $\cal V(\lambda)$.  

The quantum scenery is developed in terms of the Ambiguity function, who transforms the real-valued Wigner functions of energies $\mathbf{E}_{1}=0.5$ and $\mathbf{E}_{2}=-8$, into symmetric complex-valued functions, centered at the origin in the $\lambda-\theta$ plane, real part is shown in Fig. (\ref{fig-R_real_phase_space}) [See Appendix C, Figs. (\ref{fig-R_Imaginary_phase_space}) and (\ref{fig-R_Probability_phase_space})   to observe the imaginary part and the absolute value squared.]. Those quantum states evolve along the level sets of the conservation law given by Eq.(\ref{R_Hamiltonian-Eq}), since the generator of motion $\hat{G}$ besides to commute with $\mathcal{H}(\hat{x},\hat{p})$, also commutes with $\mathbb{H}(\hat{\lambda},\hat{\theta})$.
\begin{align}
\hat{G} = \frac{1}{m} \hat{p} \hat{\lambda} + m \omega^2 \hat{x} \hat{\theta},
\end{align}
\begin{align}
 {[} \hat{G}, \mathcal{H}(\hat{x},\hat{p})  {]} =   {[} \hat{G}, \mathbb{H}( \hat{\lambda}, \hat{\theta})  {]}  = 0.
\end{align}
This proves that $\mathcal{H}(\hat{x},\hat{p})=E$ and ${\mathbb{H}}(\hat{\lambda},\hat{\theta})={\cal E}$, are integrals of motion associated with the transformation $\widehat{U}(t)= e^{-\frac{i}{\hbar}\hat{G}t}$. Hence, the conservative dynamics established in the reciprocal phase space forbids tunneling across the $\cal V(\lambda)$ barrier, and leads us to understand the real and complex components of $A(\lambda,\theta)$ 
as probability amplitudes dragged along the well defined trajectories stated by the Eq.(\ref{R_newton_sol-Eq}). Complementarily, notice that the only completely real and positive ambiguity 
function corresponds to a Gaussian Wigner state centered at the origin in the $x-p$ plane, depicting 
a state with the highest probability to be found at the top of the IO potential barrier. 
\begin{figure}[t!]
\begin{tabular}{cc}
\label{RNewton-fig}
\includegraphics[scale=0.51,angle=0]{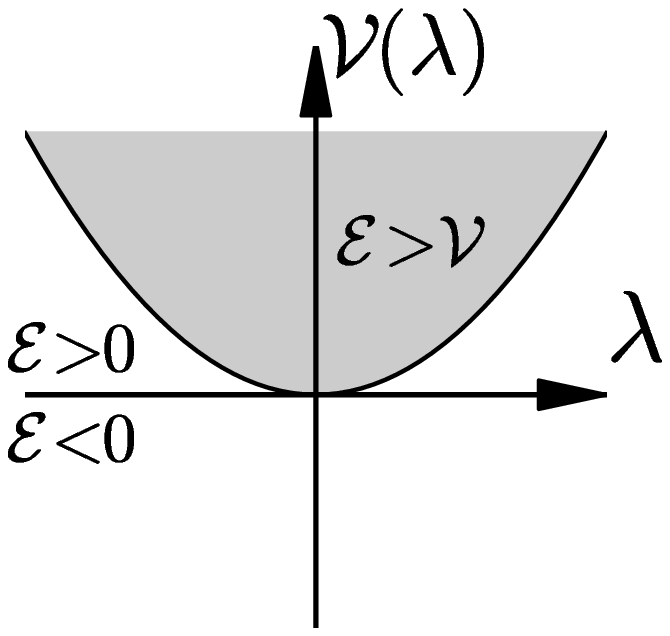} &
\includegraphics[scale=0.75,angle=0]{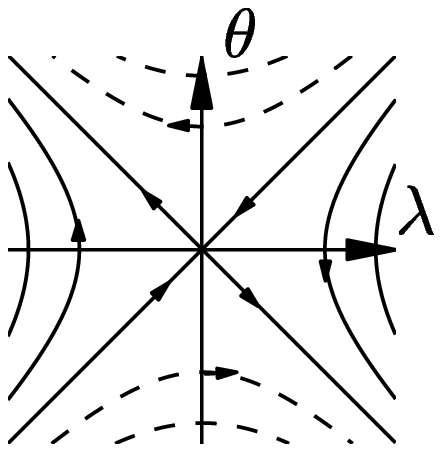} \\
(a) & (b)\\
\end{tabular}
\caption{(a) ${\cal V}(\lambda)$ barrier. (b) Reciprocal classical phase$ $-$ $space portrait: solid or dashed lines 
correspond to particles with positive or negative ${\cal E}$ respectively. The direction
of motion is represented by arrows. }
\end{figure}
\begin{figure*}[t!]
\centering
\hspace{0em}\includegraphics[scale=0.8,angle=0]{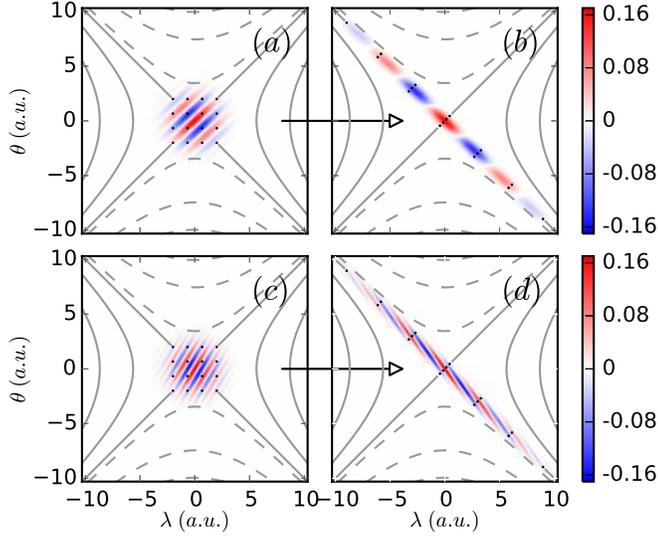}
\caption{Ambiguity function's real part $ Re[A(\lambda,\theta)]$ for energies 
$\mathbf{E}_{1}=-0.5 $ [(a)\&(b)], and $\mathbf{E}_{2}=-8 $ [(c)\&(d)] given in Wigner phase space for states subjected to the IO at times $t_{0}=0 \ a.u. $ and  $t=1.5 \ a.u.$ ($\hbar=\omega=1a.u.$). Black dots depict particles moving along the reciprocal phase space trajectories Eq.(26), and solid lines display the level set of the new conservation law Eq.(28).}
\label{fig-R_real_phase_space}
\end{figure*} 

Finally, exploiting the ambiguity function features, the doubly degenerate energy states that characterizes the IO are reflected in the reciprocal phase space by the existence of degenerate states as complex conjugates
\begin{align}
\label{A_degenerate_states}
A_{W}(\lambda,\theta) = A_{W'}^{*}(\lambda,\theta),
\end{align}
where $A_{W}$ and $A_{W'}$ denote the ambiguity functions for the pair of degenerates Wigner functions $W$ and $W'$.
\section{\small{\textbf{SUMMARY AND CONCLUSIONS}}}
We have reviewed the well-known fact that the propagation of quantum states 
subjected to quadratic Hamiltonians is described by classical equations of motion. This applies to the IO, 
for which the tunneling and reflection coefficients in phase space are given by the classically allowed phase space trajectories 
of the Wigner function, corresponding to energy components above or below the stable separatrix. 
Furthermore, this result coincides
with the path integral framework in which quadratic Hamiltonians generate exact \emph{semiclassical} propagators, 
that prevents the flow of quantum states across the classically forbidden regions. Therefore,
the \emph{propagation} of quantum states under quadratic Hamiltonias are perfectly reproduced 
by the Liouville equation, as well as the Koopman Von Neumann equation of motion that are ultimatelly 
equivalent to the propagation of Newtonian particles. 

Moreover, the most relevant contribution of this paper is the treatment 
in the reciprocal phase space that leads us to elucidate the IO as a classical dynamical
system with two conservation laws associated to the propagator $\widehat{U}(t)= e^{-\frac{i}{\hbar}\hat{G}t}$, including the energy as one of them. Despite this characteristic, a natural question is raised: whether or not there are in general cases where both quantum and classical operators share exactly the same symmetry. However, this treatment goes beyond the scope of the topic and will be subjected to further research. Moreover, another insight on the Ambiguity function is that it relates the pairs of degenerate energy states of the IO, as complex conjugates. 

In summary, for quadratic Hamiltonians, quantum dynamics can be described by the Koopman-Von Neumann equation of motion; 
determining that quantum states strictly evolve throughout classical trajectories. A similar behavior is observed in the reciprocal phase space where the state represented by the ambiguity function evolves along well defined trajectories in the reciprocal classical phase space. Finally, we stress that even if a quantum propagation is equivalent to a classical evolution, 
quantum mechanics sets additional restrictions on the states in order to maintain consistency.
\section*{\textbf{Acknowledgements}}
I would like to thank Renan Cabrera Ph.D. with whom I had long conversations regarding the subject of this paper and for the energetic  computational training. Moreover, I appreciate the opinions and recommendations expressed by A. Ticona Ph.D. and G. M. Ramirez Ph.D.


\nocite{*}
\bibliographystyle{plain}
\section*{\textbf{References}}
\bibliography{references} 

\appendix
\section*{\textbf{APPENDIX}}
\renewcommand{\thesubsection}{\Alph{subsection}}
\section{\small\textbf{HILBERT PHASE-SPACE REPRESENTATIONS}}
Schr\"odinger equation is restricted to describe closed quantum systems, {\it i.e.}, not-interacting with the environment, maintaining perfect coherence along the evolution, and entailing no-loss of information. In this formalism, the knowledge about the system is encoded on pure quantum states represented by the ket $|\psi\rangle$, who might be rewritten as a linear combination of a given complete set of eigenstates of an Hermitian operator   
\begin{align}
| \psi (t)\rangle= \sum_{n} C_{n}(t)| \phi_n \rangle,
\end{align}
the modulus squared of the coefficient $|C_{n}(t)|^2$ define the probability of finding the system in the eigenstate $| \phi_n \rangle$, then it must be true that $\sum_{n} | C_{n}(t)|^2 = 1$. This means that coherent superpositions between states are permitted, in particular a coherent superposition of two states might be constructed as
\begin{align}
 | \psi \rangle =  \frac{1}{\sqrt{2}}( | \phi_1 \rangle + | \phi_2 \rangle ).
\end{align}
An alternative representation of quantum states recast on the density operator state, constructed from a ket and its bra 
\begin{align}
\label{rho_pure-Eq}
\hat{\rho}^{pure}=\left|\psi\right\rangle \left\langle \psi \right|. %
\end{align}
Any density operator state represented on this form is called pure, because it basically contains the same information 
as the ket, up to a global phase. The advantage of the density operator state over the ket relies on it can describe statistical ensembles of pure states called mixed 
\begin{align}
\label{rho_mixed-Eq}
\hat{\rho}=\sum_{i}p_{i}\hat{\rho_{i}} =\sum_{i}p_{i} \left|\psi_{i}\right\rangle \left\langle \psi_{i} \right|, %
\end{align}
here $p_{i}$ is the probability associated to the pure state $\left|\psi_{i}\right\rangle$, on such a way that $\sum_{i}p_{i}=1$. 
Thus, incoherent superpositions who characterize mixed states are  
interpreted as a collective description of an ensemble of pure quantum states, for example
\begin{align}
 \hat{\rho} =  \frac{1}{2}(  \hat{\rho}_1 + \hat{\rho_2}   ),
\end{align}
where $\hat{\rho}_1$ and $\hat{\rho}_2$ might be constructed following Eq. (\ref{rho_pure-Eq}) 
or they can be incoherent superpositions by themselves. In summary, quantum systems that are able to be represented by a single ket are called pure states, otherwise, they are referred as mixed states.

The time evolution of the density operator state $\hat{\rho}$ is obtained by differentiating Eq.(\ref{rho_mixed-Eq}) with respect to time 
\begin{align}
\label{rho_mixed_diff-Eq}
 \dot{ \hat{\rho} } = \sum_{i}p_{i} (\dot{\left| \psi_{i} \right\rangle} \left\langle \psi_{i} \right|  + \left| \psi_{i} \right\rangle \dot{\left\langle \psi_{i} \right| } ), 
\end{align}
where $\dot{\left|\psi_{i}\right\rangle}$ and $\dot{\left\langle\psi_{i}\right|}$ are provided through Schr\"odinger equation 
\begin{align}
 \dot{\left|\psi_{i}\right\rangle} &= \frac{1}{i\hbar} \widehat{H} \left|\psi_{i}\right\rangle,\label{schr-Eq}  \\
 \dot{\left\langle\psi_{i}\right|}&= -\frac{1}{i\hbar}\widehat{H} \left\langle\psi_{i}\right|,\label{Her_schr-Eq}%
\end{align}
it turns Eq.(\ref{rho_mixed_diff-Eq}) on
\begin{align}
\
\dot{\hat{ \rho } } = \sum_{i}p_{i} (\ \ \frac{1}{i\hbar}\widehat{H} \left|\psi_{i}\right\rangle \left\langle \psi_{i} \right|  -  \frac{1}{i\hbar} \left| \psi_{i} \right\rangle \left\langle\psi_{i}\right| \widehat{H} \ \ ), %
\
\end{align}
sorting out
\begin{align}
\label{von_New_deduc-eq}
\
i\hbar \dot{\hat{ \rho} } = \widehat{H}\sum_{i}p_{i}  \left|\psi_{i}\right\rangle \left\langle \psi_{i} \right|  -  \sum_{i}p_{i} \left| \psi_{i} \right\rangle \left\langle\psi_{i}\right| \widehat{H}. %
\
\end{align}
Finally substituting Eq.(\ref{rho_mixed-Eq}) in Eq.(\ref{von_New_deduc-eq}) von Neumann equation is obtained, 
which holds for both pure and mixed states
\begin{align}
i\hbar \dot{\hat{ \rho}} = \widehat{H} \hat{\rho} -  \hat{\rho} \widehat{H}  , \label{vNeum_des-Eq} \\
i\hbar \dot{\hat{ \rho}}=[\widehat{H},\hat{\rho}]. \label{von_Neum_commutator-Eq}%
\
\end{align}
Matrix elements of the density operator state $\hat{\rho}$ in a certain basis set of kets is given by 
$\langle \boldsymbol{x} | \hat{\rho}  | \boldsymbol{y}  \rangle= \rho_{\boldsymbol{xy}}$, for example in the position representation
the above equation reads
\begin{multline}
\label{Von_Neumann_x-Eq1}
i\hbar \frac{ \partial }{\partial t}\langle \boldsymbol{x} | \rho  | \boldsymbol{x}'  \rangle = 
\Big[H \left( \boldsymbol{x},-i\hbar\frac{\partial}{\partial \boldsymbol{x} } \right) \\ 
-H \left(\boldsymbol{x}',i\hbar\frac{\partial}{\partial \boldsymbol{x}' }\right) \Big] \langle \boldsymbol{x} | \rho  | \boldsymbol{x}'  \rangle.
\end{multline}
while in the momentum representation we have
\begin{multline}
\label{voN_Neuman_P-Eq}
i\hbar \frac{ \partial }{\partial t}\langle \boldsymbol{p} | \rho  | \boldsymbol{p}'  \rangle = \Big[H \Big(i\hbar\frac{\partial}{\partial \boldsymbol{p} },\boldsymbol{p}\Big) \\ 
-H\Big(  -i\hbar\frac{\partial}{\partial \boldsymbol{p}' },\boldsymbol{p}' \Big) \Big] \langle \boldsymbol{p} | \rho  | \boldsymbol{p}'    \rangle.
\end{multline}
However, the free-coordinate formulation of von Neumann equation in the \emph{Hilbert phase space} \cite{bondar2012operational,PhysRevA.88.052108,PhysRevA.92.042122} is achieved by the 
mirror quantum operators $\boldsymbol{x}'$ and $\boldsymbol{p}'$
\begin{align}
\label{vonNeumann_appendix-Eq}
i\hbar\frac{ d }{ d t}\left|\rho\right\rangle 
=[H \left(\hat{\boldsymbol{x} } , \hat{ \boldsymbol{p}} \right) 
- H\left(\hat{\boldsymbol{x}}' , \hat{ \boldsymbol{p}}' \right)]\left|\rho\right\rangle. %
\end{align}
such that
\begin{align}
 [ \hat{\boldsymbol{x} } , \hat{\boldsymbol{p} }  ] = i \hbar \quad [ \hat{\boldsymbol{x} }' , \hat{\boldsymbol{p} }'  ] = -i \hbar,
\end{align}
\begin{align}
  [\hat{\boldsymbol{x} } , \hat{\boldsymbol{x} }'  ] = [ \hat{\boldsymbol{x} } , \hat{\boldsymbol{p}}'  ] = [ \hat{\boldsymbol{p}}' , \hat{\boldsymbol{p} }  ] =[ \hat{\boldsymbol{p} }' , \hat{\boldsymbol{x} }  ] = 0.
\end{align}
Then, the linear change of variables 
\begin{align}
  \label{eq-x-1}
 \hat{\boldsymbol{x} }  = \hat{x}-\frac{\hbar}{2}\hat{\theta}, \quad \hat{\boldsymbol{x} }' = \hat{x}+\frac{\hbar}{2}\hat{\theta},\\
  \label{eq-x-2}
  \hat{\boldsymbol{p}} =\hat{p}+\frac{\hbar}{2}\hat{\lambda} , \quad \hat{\boldsymbol{p}}'=\hat{p}-\frac{\hbar}{2}\hat{\lambda},
\end{align}
leads to
\begin{multline}
\label{vonNeumann-appendix-Eq2}
i\hbar\frac{d }{d t}\left|\rho\right\rangle 
 = [ H \left(\hat{x}-\frac{\hbar}{2}\hat{\theta} , \hat{p}+\frac{\hbar}{2}\hat{\lambda} \right) \\
- H \left( \hat{x}+\frac{\hbar}{2}\hat{\theta} ,\hat{p}-\frac{\hbar}{2}\hat{\lambda} \right)  ] \left|\rho\right\rangle. %
\end{multline}
This scheme also employs four operators ($\hat{x}$, $\hat{p}$, $\hat{\lambda},\hat{\theta}$), with the following  commuting  relations 
\cite{bondar2012operational} 
\begin{align}
[\hat{x},\hat{\lambda}]=i,   \quad [\hat{p},\hat{\theta}]=i, 
\end{align}
\begin{align}
\label{commuting_operators}
[\hat{x},\hat{p}]=0, \quad [\hat{x},\hat{\theta}]=0, \quad [\hat{\lambda},\hat{p}]=0, \quad [\hat{\lambda},\hat{\theta}]=0.
\end{align}
As a consequence, the \emph{Hilbert phase space} is parametrized by the spectrums of two commuting operators, 
selected from Eq. (\ref{commuting_operators}), due to
each pair share a common basis set of orthogonal eigenvectors: $\left|xp\right\rangle,\ \ \left|x\theta \right\rangle,\ \ \left|\lambda p\right\rangle$ and $\left|\lambda \theta \right\rangle$. In contrast, we remind to the reader that the Hilbert space requires only the definition of two non-commuting operators: $\hat{\boldsymbol{x} }$ and $\hat{\boldsymbol{p} }$, thus, it is usually parametrized 
by either $\boldsymbol{x}$ or $\boldsymbol{p}$. Thereupon, the density state operator $\hat{\rho}$ who lies in the standard Hilbert space becomes on a ket $\left|\rho(t)\right\rangle$ in the larger \emph{Hilbert phase space}. The ket $\left|\rho(t)\right\rangle$ realization is done a projection on a given base, for instance in the phase space representation [$x$-$p$] we have
$\langle xp \left|\rho(t)\right\rangle$, for a complete summary see Table \ref{spaces-tab}.

For the sake of concreteness, Von Neumann equation in the Hilbert phase space exhibits 
particular forms, depending on the representations held. They are given by the 
following expressions that are fundamentally equivalent 
\begin{itemize}
\item $x-p$ $representation$
\begin{multline}
\label{VonNeu_xp_apendice-Eq}
i\hbar\frac{\partial }{\partial t} \langle xp \left|\rho(t)\right\rangle = \Big[ H\Big(x +i \frac{\hbar}{2}\frac{\partial}{\partial p},p 
-i\frac{\hbar}{2}\frac{\partial}{\partial x}\Big) \\
- H\Big(x -i \frac{\hbar}{2}\frac{\partial}{\partial p},p +i \frac{\hbar}{2} \frac{\partial}{\partial x}\Big)\Big]\langle xp \left|\rho(t)\right\rangle. 
\end{multline}
\item $x-\theta$ $representation$
\begin{multline}
\label{VonNeu_xtheta_apendice-Eq}
\
i\hbar\frac{\partial }{\partial t} \langle x\theta \left|\rho(t)\right\rangle = \Big[ H \Big(x-\frac{\hbar}{2}     \theta,i[\frac{\partial}{\partial \theta}-\frac{\hbar}{2}\frac{\partial}{\partial x}]\Big) \\
- H\Big(x+\frac{\hbar}{2}\theta,i[\frac{\partial}{\partial \theta}+ \frac{\hbar}{2}\frac{\partial}{\partial x}]\Big)\Big] \langle x\theta \left|\rho(t)\right\rangle. 
\
\end{multline}
\item $\lambda-p$ $representation$
\begin{multline}
\label{VonNeu_lambdap_apendice-Eq}
\
i\hbar\frac{\partial }{\partial t} \langle \lambda p \left|\rho(t)\right\rangle = \Big[ H\Big(i[\frac{\partial}{\partial \lambda}+\frac{\hbar}{2}\frac{\partial}{\partial p}],p+\frac{\hbar}{2}\lambda\Big) \\
- H\Big(i[\frac{\partial}{\partial \lambda}-\frac{\hbar}{2}\frac{\partial}{\partial p}],p-\frac{\hbar}{2}\lambda\Big)\Big] \langle \lambda p \left|\rho(t)\right\rangle. 
\
\end{multline}
\item $\lambda-\theta$ $representation$
\begin{multline}
\label{VonNeu_lambda_theta_apendice-Eq}
\
i\hbar\frac{\partial }{\partial t} \langle \lambda \theta \left|\rho(t)\right\rangle = \Big[ H\Big( i\frac{\partial}{\partial \lambda}-\frac{\hbar}{2}\theta,i\frac{\partial}{\partial \theta}+\frac{\hbar}{2}\lambda\Big) \\
- H\Big( i\frac{\partial}{\partial \lambda}+\frac{\hbar}{2}\theta,i\frac{\partial}{\partial \theta}-\frac{\hbar}{2}\lambda\Big)\Big] \langle \lambda \theta \left|\rho(t)\right\rangle. 
\
\end{multline}
\end{itemize}
\begin{table*}[pt!]
\caption{\label{spaces-tab}  This table shows all representations hold by the Hilbert phase space. The explicit form for the extended four operators is constructed such that the commutation relations given by the Eq.(\ref{extended-commutators}) are fulfilled.} 
\begin{tabular}{ccccc}
\hline\hline
 Space& Main commuting &Basis&Completeness & Extended four operator algebra \\
 & relation & & identity  &   \\ \hline \\
 $xp-representation$&$[\hat{x},\hat{p}]=0$&$\left|xp \right\rangle$&$1 =\int dxdp \left|xp\right\rangle \left\langle xp\right|$&$\hat{x}=x,   \quad  \hat{p}=p, \quad  \hat{\lambda}=-i\frac{\partial} {\partial x}, \quad   \hat{\theta}
 =-i \frac{\partial}{\partial p}$ \\ \\ \hline \\
 $x\theta-representation$&$[\hat{x},\hat{\theta}]=0$
 &$\left|x\theta \right\rangle$&$1 =\int dxd\theta \left|x\theta\right\rangle \left\langle x\theta\right|$&$\hat{x}=x,  \quad    \hat{p}=i\frac{\partial} {\partial \theta},  \quad  \hat{\lambda}=-i\frac{\partial} {\partial x}, \quad    \hat{\theta}=\theta.$\\ \\ \hline \\
 $\lambda p-representation$&$[\hat{\lambda},\hat{p}]=0$&$\left|\lambda p\right\rangle$&$1 =\int d\lambda dp \left|\lambda p\right\rangle \left\langle \lambda p\right|$&$\hat{x}=i\frac{\partial} {\partial \lambda},   \quad   \hat{p}=p,  \quad  \hat{\lambda}=\lambda, \quad    
\hat{\theta}=-i\frac{\partial} {\partial p}.$\\ \\ \hline \\
 $\lambda\theta-representation$&$[\hat{\lambda},\hat{\theta}]=0$&$\left|\lambda \theta \right\rangle$&$1 =\int d\lambda d\theta \left|\lambda \theta\right\rangle \left\langle \lambda \theta\right| $&$\hat{x}=i\frac{\partial} {\partial \lambda},  \quad  \hat{p}=i\frac{\partial} {\partial \theta},  \quad  \hat{\lambda}=\lambda, \quad   \hat{\theta}=\theta.$\\ \\
\hline\hline
\end{tabular}
\footnote{}\\
${}^1$\footnotesize {where ${\small \left\langle \lambda p\right|x\theta\rangle=\mathrm{exp}(ip\theta -ix\lambda)/(2\pi)}$.}
\end{table*}
\section{\small\textbf{THE WIGNER FUNCTION AND OTHER REPRESENTATIONS OF QUANTUM STATES}}
In this appendix we will derive the connection among well-known distributions functions employed to represent a quantum state, and how they are related to the ket $\left|\rho(t)\right\rangle$ representations hold by \emph{the Hilbert phase space}.
According to the parametrization, there are four distributions functions in consideration: 
the Wigner function $W(x,p;t)$ \cite{wigner1932quantum}, the double 
configuration space representation $B(x,\theta;t)$ \cite{Blokhintsev1940a,Blokhintsev1941} introduced by Blokhintsev \cite{kuzemsky2008works}, 
the double-momentum-space representation $Z(\lambda,p;t)$ and the Ambiguity function $A(\lambda,\theta;t)$ \cite{cohen1985generalized,cohen1989time}. 

The ``Double-configuration-space-function'' or Blokhintsev function is defined as
\begin{align}
B(x,\theta;t) = \langle x - \frac{\hbar}{2} \theta | \rho(t)  |x + \frac{\hbar}{2} \theta   \rangle,
\end{align}
and more precisely for pure states is reduced to 
\begin{align}
\label{B_pure-Eq}
B(x,\theta;t)=\psi(x-\frac{\hbar}{2}\theta; t)\psi^*(x+\frac{\hbar}{2}\theta;t). %
\end{align}
The motion equation for the Blokhintsev function, $B(x,\theta;t)$ is
\begin{multline}
\label{B_motion-Eq}
 i\hbar\frac{\partial}{\partial t}B(x,\theta;t) = \Big[ H\Big(x-\frac{\hbar}{2}     \theta,i[\frac{\partial}{\partial \theta}-\frac{\hbar}{2}\frac{\partial}{\partial x}]\Big) \\
- H\Big(x+\frac{\hbar}{2}\theta,i[\frac{\partial}{\partial \theta}+ \frac{\hbar}{2}\frac{\partial}{\partial x}]\Big)\Big]B(x,\theta;t).
\end{multline}
Then, Wigner function might be obtained through a inverse Fourier transform on $B(x,\theta;t)$
\begin{align}
\label{Wigner-transform-eq}
W(x,p;t)= \frac{1}{2\pi}\int B(x, \theta;t) e^{ip\theta}d\theta.
\end{align}
Wigner function's motion equation is named Moyal's equation, in honor to the physicist Jos\'e Enrique Moyal (1910-1998) 
\begin{multline}
\label{Moyal-Eq}
i\hbar \frac{\partial{W(x,p;t)}}{\partial{t}} = \Big[H\Big(x +i \frac{\hbar}{2}\frac{\partial}{\partial p},p 
-i\frac{\hbar}{2}\frac{\partial}{\partial x}\Big) \\
- H\Big(x -i \frac{\hbar}{2}\frac{\partial}{\partial p},p +i \frac{\hbar}{2} \frac{\partial}{\partial x}\Big) \Big]W(x,p;t). 
\end{multline}
Applying a Fourier transform on the Wigner function we get the Double-momentum-space representation $Z(\lambda,p;t)$, who name is owed 
to the fact that $\hbar\lambda$ has the dimension of momentum
\begin{align} 
\label{Z-transform-eq}
Z(\lambda,p;t)= \int W(x, p;t) e^{-ix\lambda}dx,
\end{align}
obeying
\begin{multline}
\label{Z-dyn-Eq}
i\hbar\frac{\partial }{\partial t} Z(\lambda,p;t)  = \Big[ H\Big(i[\frac{\partial}{\partial \lambda}+\frac{\hbar}{2}\frac{\partial}{\partial p}],p+\frac{\hbar}{2}\lambda\Big) \\
- H\Big(i[\frac{\partial}{\partial \lambda}-\frac{\hbar}{2}\frac{\partial}{\partial p}],p-\frac{\hbar}{2}\lambda\Big)\Big]Z(\lambda,p;t) . 
\end{multline}
In brief, the connection among these functions are obtained through partial Fourier transforms, keeping in mind that $\lambda$ is the conjugate variable of $x$, and $\theta$ is the conjugate variable of $p$ 
\begin{align}
\lambda   \overset{\mathcal{F}}{\rightarrow} x, \quad  \lambda  \overset{\mathcal{F}^{-1}}{\rightarrow} x, \\
\theta \overset{\mathcal{F}}{\rightarrow} p,   \quad \theta \overset{\mathcal{F}^{-1}}{\rightarrow}  p.
\end{align}
For instance, the Ambiguity function 
$A(\lambda,\theta;t)$ given by Eq.(\ref{A-eq-1}) can be alternatively obtained by
\begin{align}
\label{A-eq-2}
A(\lambda,\theta;t)= \int B(x, \theta;t) e^{-i\lambda x}dx,
\end{align}
or
\begin{align}
\label{A-eq-3}
A(\lambda,\theta;t)= \int Z(\lambda, p;t) e^{-ip \theta}dp.
\end{align}
The ambiguity function obeys the following motion equation
\begin{multline}
\label{A_motion-Eq}
i\hbar\frac{\partial}{\partial t}A(\lambda, \theta;t) = \Big[ H\Big( i\frac{\partial}{\partial \lambda}-\frac{\hbar}{2}\theta,i\frac{\partial}{\partial \theta}+\frac{\hbar}{2}\lambda\Big) \\
- H\Big( i\frac{\partial}{\partial \lambda}+\frac{\hbar}{2}\theta,i\frac{\partial}{\partial \theta}-\frac{\hbar}{2}\lambda\Big)\Big]A(\lambda,\theta;t).
\end{multline}
In addition, the four distributions functions presented are proportional 
to the different representations of the ket in the \emph{Hilbert phase space}, through  
\begin{align}
\label{Relations-Hilb-BW-Eq}
B(x,\theta;t) =  \frac{1}{\sqrt{ \hbar}} \langle  x\theta | \rho(t)  \rangle,  \quad  Z(\lambda,p;t) =  \frac{1}{\sqrt{ \hbar}} \langle  \lambda p | \rho (t) \rangle,
\end{align}
\begin{align}
\label{Relations-Hilb-ZA-Eq}
W(x,p;t) =  \frac{1}{\sqrt{2\pi \hbar}} \langle  xp | \rho(t)  \rangle,  \quad  A(\lambda,\theta;t) =  \frac{1}{\sqrt{\hbar}} \langle  \lambda    
\theta | \rho(t)  \rangle.
\end{align}
The first equivalence is established by writing the completeness identity for the quantum observables $\boldsymbol{x}$ and $\boldsymbol{x'} $
\begin{align}
\label{completenessXX'-Eq}
1 =\int d \boldsymbol{x} d \boldsymbol{x}' \left| \boldsymbol{x} \boldsymbol{x}'\right\rangle \left\langle \boldsymbol{x} \boldsymbol{x}'\right|,
\end{align}
then the spatial linear change given by Eq.(\ref{eq-x-1}) in the $x-\theta$ representation is applied 
\begin{align}
1 =\int  | \boldsymbol{J}(\boldsymbol{x},\boldsymbol{x}') | dx d\theta  \left| x - \frac{\hbar}{2}\theta ,x + \frac{\hbar}{2}\theta  \right\rangle \left\langle  x - \frac{\hbar}{2}\theta ,x + \frac{\hbar}{2}\theta  \right| , 
\end{align}
with 
\begin{align}
| \boldsymbol{J}(\boldsymbol{x},\boldsymbol{x}') | = \begin{vmatrix}
\frac{\partial \boldsymbol{x} } {\partial x } & \frac{\partial \boldsymbol{x} } {\partial \theta } \\ 
 \frac{\partial \boldsymbol{x}' } {\partial x } & \frac{\partial \boldsymbol{x}' } {\partial \theta }
\end{vmatrix} = \begin{vmatrix}
1 & -\frac{\hbar}{2} \\ 
1 & \frac{\hbar}{2} 
\end{vmatrix} 
=\hbar,
\end{align}
thus,
\begin{align}
\label{completenessXTHETA-2_Eq}
1 =\int  \hbar  dx d\theta  \left| x - \frac{\hbar}{2}\theta ,x + \frac{\hbar}{2}\theta  \right\rangle \left\langle  x - \frac{\hbar}{2}\theta ,x + \frac{\hbar}{2}\theta  \right|.  
\end{align}
Moreover, from the completeness identity for the $x$ and $\theta$ quantum observables we have
\begin{align}
\label{completenessXTHETA-Eq}
1 =\int dxd\theta \left|x\theta\right\rangle \left\langle x\theta\right|.
\end{align}
It follows that Eqs.(\ref{completenessXTHETA-2_Eq}) and (\ref{completenessXTHETA-Eq}) enable to deduce
\begin{align}
|x \theta \rangle = \sqrt{\hbar} | x - \frac{\hbar}{2}\theta, x + \frac{\hbar}{2}\theta \rangle,
\end{align}
hence, the ket $|\rho(t) \rangle$ projection on the above basis leads
\begin{multline}
\langle  x \theta | \rho(t)  \rangle = \sqrt{\hbar} \langle   x - \frac{\hbar}{2}\theta, x + \frac{\hbar}{2}\theta | \rho (t)  \rangle = \\
\sqrt{\hbar} \langle   x - \frac{\hbar}{2}\theta | \rho(t) | x + \frac{\hbar}{2}\theta \rangle,  
\end{multline}
where $B(x,\theta;t) $ is easily recognized
\begin{align}
\langle x \theta | \rho(t)  \rangle = \sqrt{\hbar} \langle   x - \frac{\hbar}{2}\theta | \rho(t) | x + \frac{\hbar}{2}\theta \rangle = \sqrt{\hbar} B(x,\theta;t), 
\end{align}
\begin{align}
B(x,\theta;t) = \frac{1}{\sqrt{\hbar}} \langle  x \theta | \rho(t)  \rangle.
\end{align}
The second relation is deduced from 
\begin{align}
1 =\int d \boldsymbol{p} d \boldsymbol{p}' \left| \boldsymbol{p} \boldsymbol{p}'\right\rangle \left\langle \boldsymbol{p} \boldsymbol{p}'\right|,
\end{align}
thereupon the linear change of variable given by Eq.(\ref{eq-x-1}) in the $\lambda-p$ representation, brought the above relation to
\begin{align}
\label{completenessPP-eq}
1 =\int  | \boldsymbol{J}(\boldsymbol{p},\boldsymbol{p}') | d\lambda dp  \left| p + \frac{\hbar}{2}\lambda ,p - \frac{\hbar}{2}\lambda  \right\rangle \left\langle  p + \frac{\hbar}{2}\lambda ,p - \frac{\hbar}{2}\lambda  \right|,
\end{align}
\begin{align}
| \boldsymbol{J}(\boldsymbol{p},\boldsymbol{p}') | =
\begin{vmatrix}
 \frac{\partial \boldsymbol{p} } {\partial \lambda } & \frac{\partial \boldsymbol{p} } {\partial p } \\ 
 \frac{\partial \boldsymbol{p}' } {\partial \lambda } & \frac{\partial \boldsymbol{p}' } {\partial p }
\end{vmatrix} = \begin{vmatrix}
\frac{\hbar}{2} & 1 \\ 
-\frac{\hbar}{2} & 1
\end{vmatrix}
= \hbar,
\end{align}
or
\begin{align}
\label{completeness-2-1}
1 =\int \hbar d\lambda dp  \left| p + \frac{\hbar}{2}\lambda ,p - \frac{\hbar}{2}\lambda  \right\rangle \left\langle p + \frac{\hbar}{2}\lambda ,p - \frac{\hbar}{2}\lambda   \right|, 
\end{align}
and from the completeness identity for $x$ and $\theta$ 
\begin{align}
\label{completeness-2-2}
1 =\int d\lambda dp \left|\lambda p\right\rangle \left\langle \lambda p\right| .
\end{align}
Then, the $\left|\lambda p\right\rangle$ basis is identified from a comparison 
between Eqs.(\ref{completeness-2-1}) and (\ref{completeness-2-2})
\begin{align}
\left|\lambda p\right\rangle = \sqrt{\hbar} |p + \frac{\hbar}{2}\lambda ,p - \frac{\hbar}{2}\lambda \rangle,
\end{align}
therefore, 
\begin{multline}
 \langle  \lambda p | \rho (t) \rangle = \sqrt{\hbar} \langle   p + \frac{\hbar}{2}\lambda, p - \frac{\hbar}{2}\lambda | \rho(t)  \rangle \\
= \sqrt{\hbar} \langle   p + \frac{\hbar}{2}\lambda  | \rho(t) | p - \frac{\hbar}{2}\lambda \rangle = \sqrt{\hbar} Z(\lambda,p;t), 
\end{multline}
\begin{align}
Z(\lambda,p;t) = \frac{1}{\sqrt{\hbar}} \langle  \lambda p | \rho (t) \rangle.
\end{align}
Finally, the remaining equivalences are easily proved by means of partial Fourier transforms.

\section{\small\textbf{REPRESENTATIONS FOR THE IO}}
Having deduced Moyal's equation in Eq.(\ref{Moyal-Eq}) it is straightforward to obtain the particular form of the motion equation for the IO Hamiltonian 
Eq.(\ref{IO-hamil-Eq}), as follows
\begin{multline}
\label{deduccion_IO_moyal-Eq}
i\hbar\frac{\partial}{\partial t}W(x,p;t) = \Big[ \frac{1}{2m} \Big( p 
-i\frac{\hbar}{2}\frac{\partial}{\partial x} \Big)^{2}-\frac{1}{2} m \omega^2 \Big( x +i \frac{\hbar}{2}\frac{\partial}{\partial p} \Big)^{2} \\
 -\frac{1}{2m} \Big( p +i \frac{\hbar}{2} \frac{\partial}{\partial x} \Big)^{2}+\frac{1}{2} m \omega^2 \Big( x -i \frac{\hbar}{2}\frac{\partial}{\partial p} \Big)^{2} \Big]W(x,p;t).
\end{multline}
Upon expanding and simplifying we arrive to Eq.(\ref{Moyal_IHO-Eq}). On the other hand, incorporating the IO 
Hamiltonian into Eq.(\ref{A_motion-Eq}), after simplification Eq.(\ref{Motion_Ambiguity_IO-Eq}) is obtained
\begin{multline}
\label{VonNeu_lambda_theta_apendice-Eq2}
i\hbar\frac{\partial}{\partial t}A(\lambda, \theta;t) = \Big[ \frac{1}{2m} \Big( i\frac{\partial}{\partial \theta}+\frac{\hbar}{2}\lambda\Big)^{2}-\frac{1}{2} m \omega^2 \Big( i\frac{\partial}{\partial \lambda}-\frac{\hbar}{2}\theta \Big)^{2} \\
 -\frac{1}{2m} \Big( i\frac{\partial}{\partial \theta}-\frac{\hbar}{2}\lambda\Big)^{2}+\frac{1}{2} m \omega^2 \Big( i\frac{\partial}{\partial \lambda}+\frac{\hbar}{2}\theta \Big)^{2} \Big]A(\lambda,\theta;t).
\end{multline}

It is important to pointed out that the phase space and the reciprocal phase space are the only representations 
for which the IO is exactly solvable. 
In contrast, the complexity in the $x-\theta $ and $\lambda-p$ representations 
require the use of second order partial differential equations, as show below
\begin{align}
\label{Motion_B_IO-Eq}
\frac{\partial B (x,\theta;t)}{\partial t} =\left[ \frac{1}{m} \frac{\partial^2}{\partial \theta \partial x} + m {\omega}^2 x \theta  \right] B(x,\theta;t),\\
\label{Motion_Z_IO-Eq}
\frac{\partial }{\partial t}Z(\lambda,p;t)
 =\left[ \frac{p\lambda }{m}   + m {\omega}^2 \frac{\partial^2}{\partial \lambda \partial p}  \right]Z(\lambda,p;t).
\end{align}
\begin{figure*}[t!]
\centering
\hspace{0em}\includegraphics[scale=0.8,angle=0]{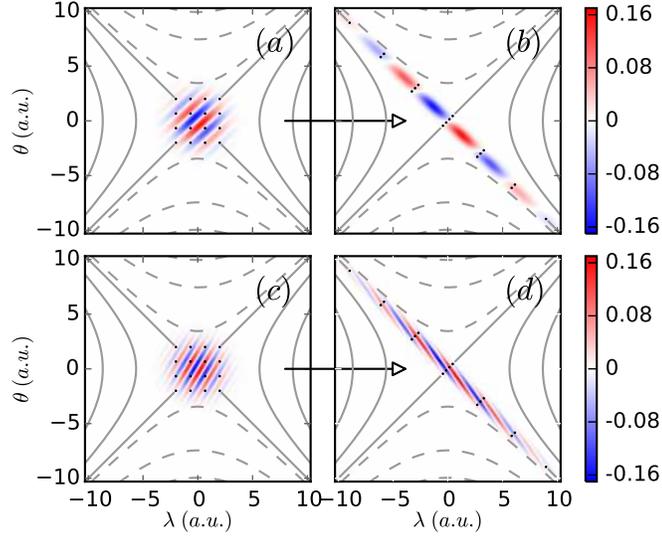}
\caption{Ambiguity function's imaginary part $ Im[A(\lambda,\theta)]$ for energies $\mathbf{E}_{1}=-0.5 $ [(a)\&(b)], and $\mathbf{E}_{2}=-8 $ [(c)\&(d)] given in Wigner phase space for states subjected to the IO at times $t_{0}=0 \ a.u. $ and  $t=1.5 \ a.u.$ ($\hbar=\omega= 1a.u.$). Black dots depict particles moving along the reciprocal phase space trajectories Eq.(26), and solid lines display the level set of the new conservation law Eq.(28).  }
\label{fig-R_Imaginary_phase_space}
\end{figure*}
\begin{figure*}[t!]
\centering
\hspace{0em}\includegraphics[scale=0.8,angle=0]{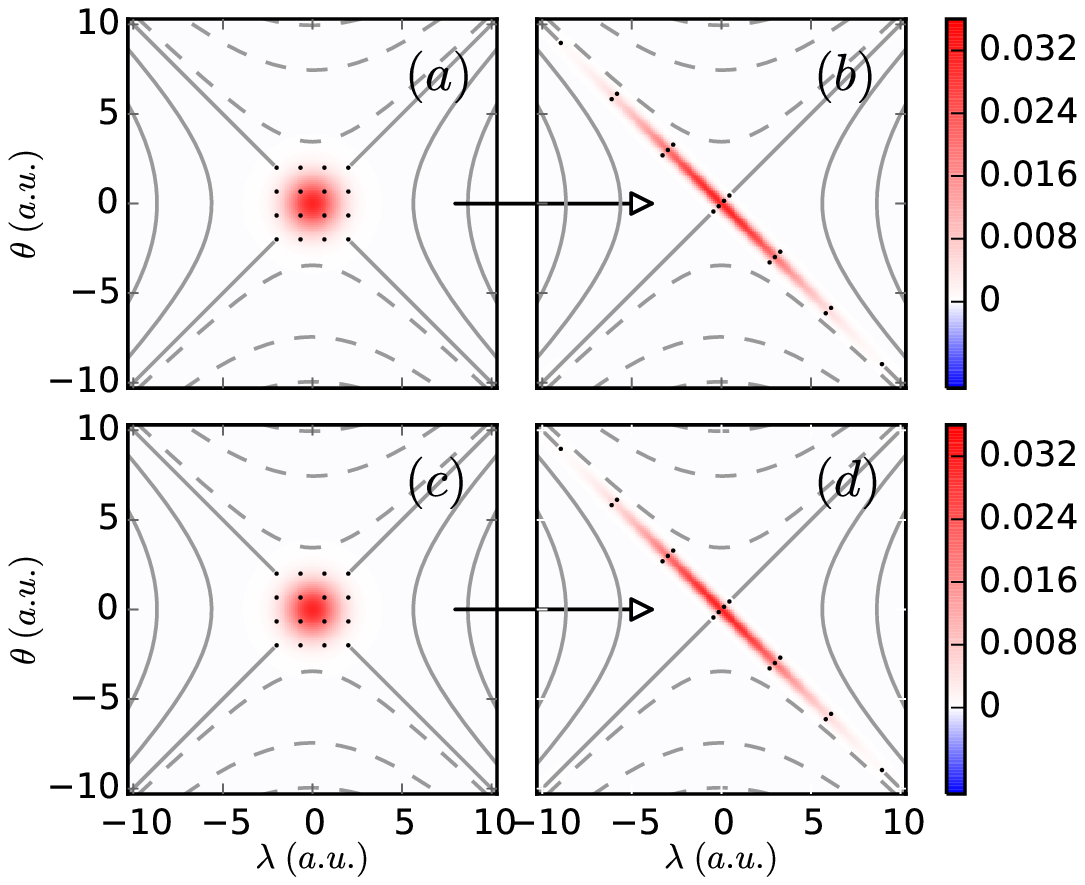}
\caption{Ambiguity function's absolute value square $ \mid A(\lambda,\theta) \mid ^2 $ for energies $\mathbf{E}_{1}=-0.5 $ [(a)\&(b)], and $\mathbf{E}_{2}=-8 $ [(c)\&(d)] given in Wigner phase space for states subjected to the IO at times $t_{0}=0 \ a.u. $ and  $t=1.5 \ a.u.$ ($\hbar=\omega= 1a.u.$). Black dots depict particles moving along the reciprocal phase space trajectories Eq.(26), and solid lines display the level set of the new conservation law Eq.(28).}
\label{fig-R_Probability_phase_space}
\end{figure*}

\end{document}